\documentclass[aps, pra, reprint, superscriptaddress]{revtex4-2}

\usepackage{xcolor}

\usepackage[letterpaper,top=2cm,bottom=2cm,left=3cm,right=3cm,marginparwidth=1.75cm]{geometry}

\definecolor{zwcolor}{rgb}{0.5,0,0.5}

\definecolor{mjocolor}{rgb}{0,0.5,0.5}

\usepackage{amsmath}
\usepackage{graphicx}
\usepackage{booktabs}
\usepackage[colorlinks=true, allcolors=blue]{hyperref}

\begin{document}
\title{Quantum Resources Required for Binding Affinity Calculations of Amyloid beta}

\author{Matthew Otten}
\affiliation{Department of Physics, University of Wisconsin – Madison, Madison, WI, USA}
\affiliation{Corresponding author: mjotten@wisc.edu}
\author{Thomas W. Watts}
\affiliation{HRL Laboratories, LLC, Malibu, CA, USA}
\author{Samuel D. Johnson}
\affiliation{HRL Laboratories, LLC, Malibu, CA, USA}
\author{Rashmi Sundareswara}
\affiliation{HRL Laboratories, LLC, Malibu, CA, USA}
\author{Zhihui Wang}
  \affiliation{Quantum Artificial Intelligence Laboratory (QuAIL), NASA Ames Research Center, Moffett Field, CA}
  \affiliation{Research Institute for Advanced Computer Science (RIACS), USRA, Moffett Field, CA}
\author{Tarini S. Hardikar}
\affiliation{qBraid Co., 111 S Wacker Dr., Chicago, IL 60606, USA}
\author{Kenneth Heitritter}
\affiliation{qBraid Co., 111 S Wacker Dr., Chicago, IL 60606, USA}
\author{James Brown}
\affiliation{qBraid Co., 111 S Wacker Dr., Chicago, IL 60606, USA}\
\author{Kanav Setia}
\affiliation{qBraid Co., 111 S Wacker Dr., Chicago, IL 60606, USA}
\author{Adam Holmes}
\affiliation{HRL Laboratories, LLC, Malibu, CA, USA}
\affiliation{Corresponding author: aholmes@hrl.com}
\begin{abstract}
\noindent Amyloid beta, an intrinsically disordered protein, plays a seemingly important but not well-understood role in neurodegenerative diseases like Alzheimer's disease. A key feature of amyloid beta, which could lead to potential therapeutic intervention pathways, is its binding affinity to certain metal centers, like iron and copper. Numerically calculating such binding affinities is a computationally challenging task, involving strongly correlated metal centers. A key bottleneck in understanding the binding affinity is obtaining estimates of the ground state energy. Quantum computers have the potential to accelerate such calculations but it is important to understand the quantum resources required. In this work, we detail a computational workflow for binding affinity calculations for amyloid beta utilizing quantum algorithms, providing estimated quantum resources required, at both the logical and hardware level.
\end{abstract}

\maketitle

\section{Introduction}
Proteins which contain a metal ion cofactor, known as metalloproteins, account for nearly half of all proteins in 
nature, with functions ranging from photosynthesis to nitrogen fixation to important biological 
functions~\cite{metalloprotein}. Metal protein interactions represent a strongly correlated system, where 
standard classical electronic structure techniques fail to accurately predict features such as coordination and 
dynamics~\cite{lloyd1996universal}. Quantum computers have long been predicted to be able to solve such strongly correlated 
problems exponentially faster than classical computers~\cite{cao2019quantum}. 
In fact, nitrogenase, which serves as
an important protein in natural nitrogen fixation, 
through its primary cofactor, iron molydbenum cofactor (FeMo-co), has long served as an example of a problem
that future, fault-tolerant quantum computers may be able to solve
and is often used as a basis for quantum resource estimation~~\cite{reiher2017elucidating,li2019electronic,otten2023qrechem}.

In their biological function, certain metalloproteins, specifically
the protein amyloid beta (A$\beta$) interacting with metal ions such as copper, ion, and zinc, have been
linked to neurodegenerative diseases such as Alzheimer's disease~\cite{warmlander_metal_2019}. These interactions
have been studied via many classical techniques, including molecular dynamics (MD), quantum mechanics / 
molecular mechanics (QM/MM), and density functional theory (DFT)~\cite{StrodelReview}, but a lack of 
accuracy has resulted in conflicting predictions for coordination schemes and other chemical quantities.
Higher accuracy calculations are therefore necessary to fully understand A$\beta$'s role in 
the onset of neurodegenerative diseases. In this paper, we provide specific instances of A$\beta$-metal ion
systems and further provide estimates of the required quantum resources to perform accurate calculations
on such systems. 

\begin{figure}
    \centering
    \includegraphics[width=0.45\textwidth]{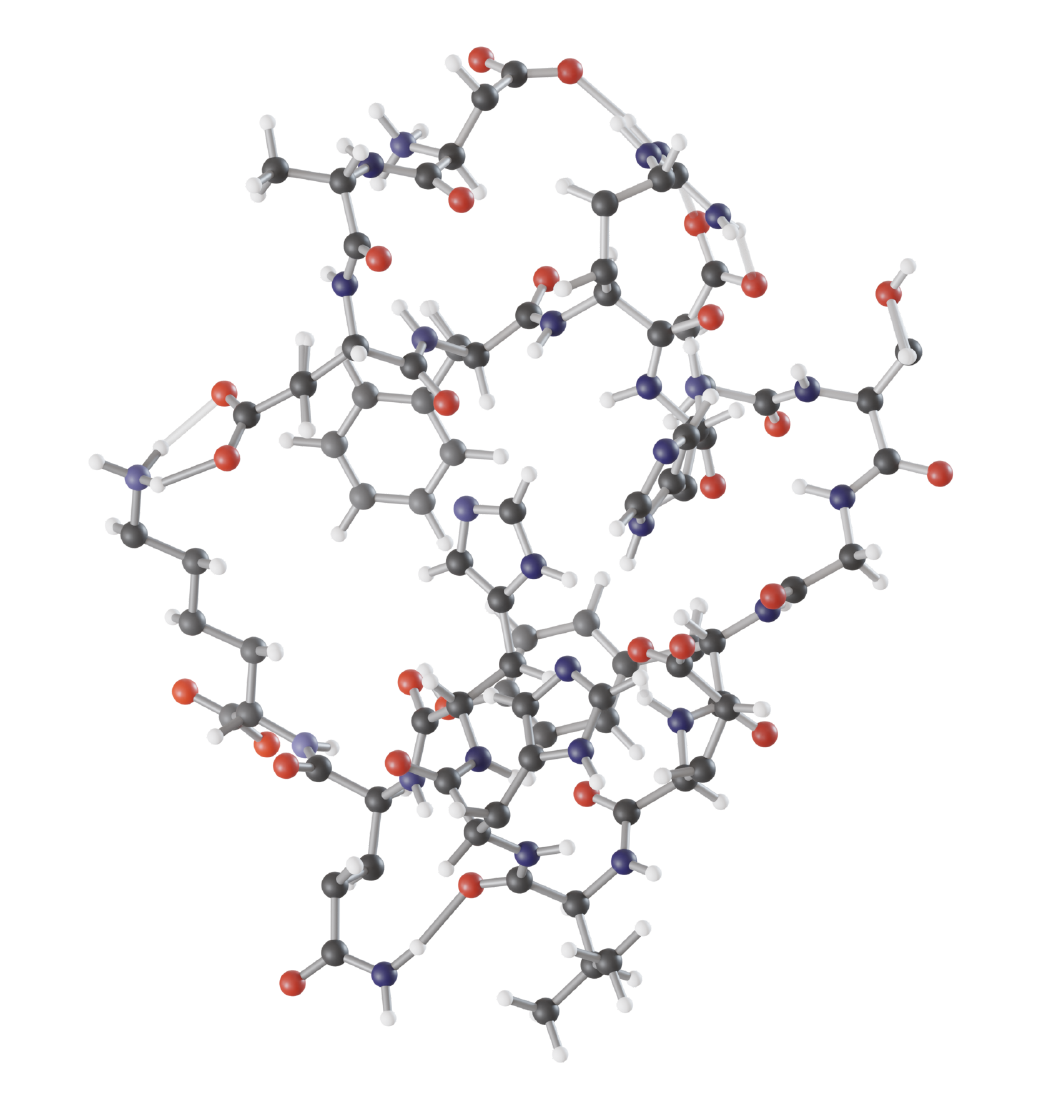}
    \includegraphics[width=0.45\textwidth]{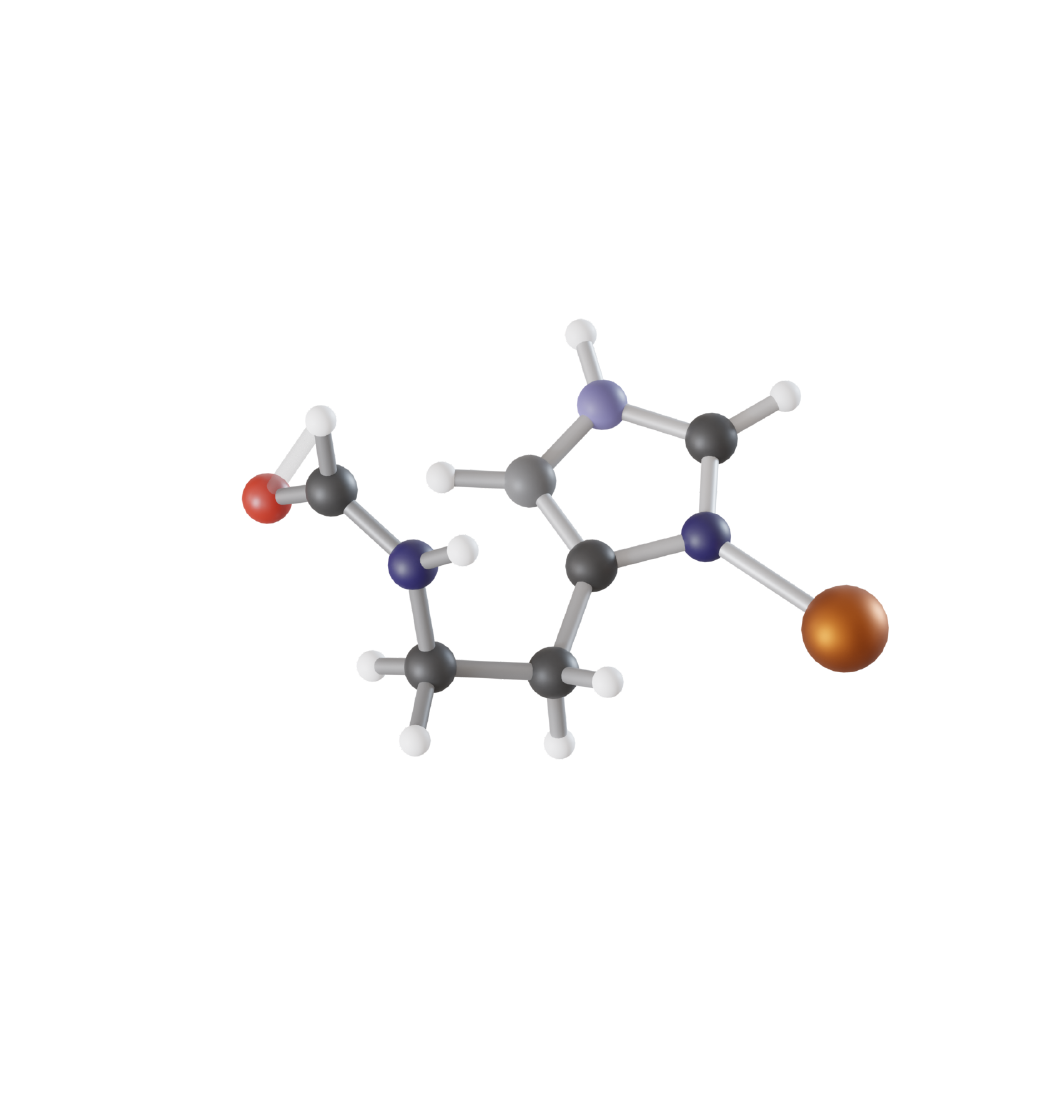}
    \caption{Top: The full structure of the AB16 protein. Bottom: A possible binding site (His6) with a Cu metal ion.}
    \label{fig:protein}
\end{figure}
\section{Problem Statement and Utility}
Amyloid-beta is a protein that is critical to understanding the pathogenesis of Alzheimer's Disease (AD). 
Created by proteases breaking down Amyloid Precursor Protein (APP), Amyloid-Beta (A$\beta$) is a disordered
protein and the subject of many leading AD hypotheses. Chief among these hypotheses is the amyloid cascade 
hypothesis and the oligomer cascade hypothesis (are soluble low molecular weight oligomers causing toxicity 
or is it the abnormal accumulation of plaques in the brain? \cite{daggetthypotheses}). To understand this 
aggregation and to design targeted drugs, a key challenge is understanding how A$\beta$ interacts with metal ions 
such as zinc, copper, iron, and platinum \cite{Metal1, Metal2}. 

Therefore, it is important to correctly model these metalloproteins and calculate their metal binding affinities.
There are many possible metal-binding domains that could be studied; AB16 (shown in Fig.~\ref{fig:protein}) is 
considered the minimal metal-binding domain for the larger A$\beta$ protein (Protein Databank (PDB) ID: 1ZE9) \cite{1ZE9}. At
physiologically relevant conditions (pH 6.5), many possible different binding sites are found for AB16.
Nuclear magnetic resonance (NMR) studies support His6, His13, His14, and Glu11 as the binding sites \cite{1ZE9};
one such region, His6, is shown interacting with a copper ion in Fig.~\ref{fig:protein}.
Experimental studies are faced with the challenges of highly disordered structure, the rapid kinetics of 
aggregation, and solvent effects when identifying coordination schemes, demonstrating the need for computational
tools.  Computational studies have been performed on AB16, with a variety of techniques used, such as QM/MM, 
classical MD, and DFT. Each approach has its own shortcomings, and often suggests alternative coordination 
schemes such as with oxygen, COAla2, Tyr10, Asp1, N-terminal nitrogen, or even water as the fourth coordination 
site~\cite{StrodelReview}. The specific computational task is to calculate the metal binding affinity of the
AB16 protein. The specific process for calculating this is detailed in the workflow below. Knowing the 
metal binding affinity provides insight into how AB16 interacts with metal ions and can be used to 
inform a broader theory to understand its role in plaque aggregation and to potentially design targeted drugs. For 
instance, a correct understanding of metal-protein coordination sites can allow us to understand the mechanism and 
kinetics of protein oligomerization and aggregation, show what ions contribute to this effect, and what 
physiological factors are necessary for this process to occur. A binding energy based mechanistic understanding 
can be the key to studying one of the earliest points of AD diagnosis. This, in turn, can be useful in future drug 
discovery and design pipelines, which could help alleviate the burden of AD. The specific problem we study, that 
of the AB16 protein, is only one in a family of possible A$\beta$ 
metalloproteins. Larger proteins (such as AB40 or AB42) could also be potential computational targets. 
The larger set of metalloproteins also include systems important for photosynthesis, nitrogen fixation, and
water oxidation~\cite{metalloprotein}. The techniques described here for AB16 could be applied to the wider
family of metalloproteins.

Alzheimer's disease has a large economic impact; the Alzheimer's Association reported that in 2024 the total 
payments for health care, long-term care and hospice services for people age 65 and older with dementia was 
estimated to be \$360 billion~\cite{alz_total_burden}. It is certainly not expected that calculations of 
the metal-binding affinity of AB16 will directly lead to a therapeutic for AD. To attempt to quantify the potential
utility of a successful computational solution for the metal-binding affinity in A$\beta$, we instead look
at the National Institutes of Health (NIH) RePORTER tool, which reports, among other things, the 
research expenditures of the NIH in an open, searchable manner~\cite{NIHRePORTERtool}. Since 2005, the NIH has 
spent around \$8.9 billion on awards for projects mentioning `amyloid beta' (and other variants, such
as `beta amyloid'). From that set of awards, a total of around \$280 million mention the word 
`computational' in the abstract. Therefore, we estimate that a technique or device which could solve
the computational problems detailed in this paper would have a utility of at least \$280 million.

\begin{figure}
    \centering
    \includegraphics[width=0.45\textwidth]{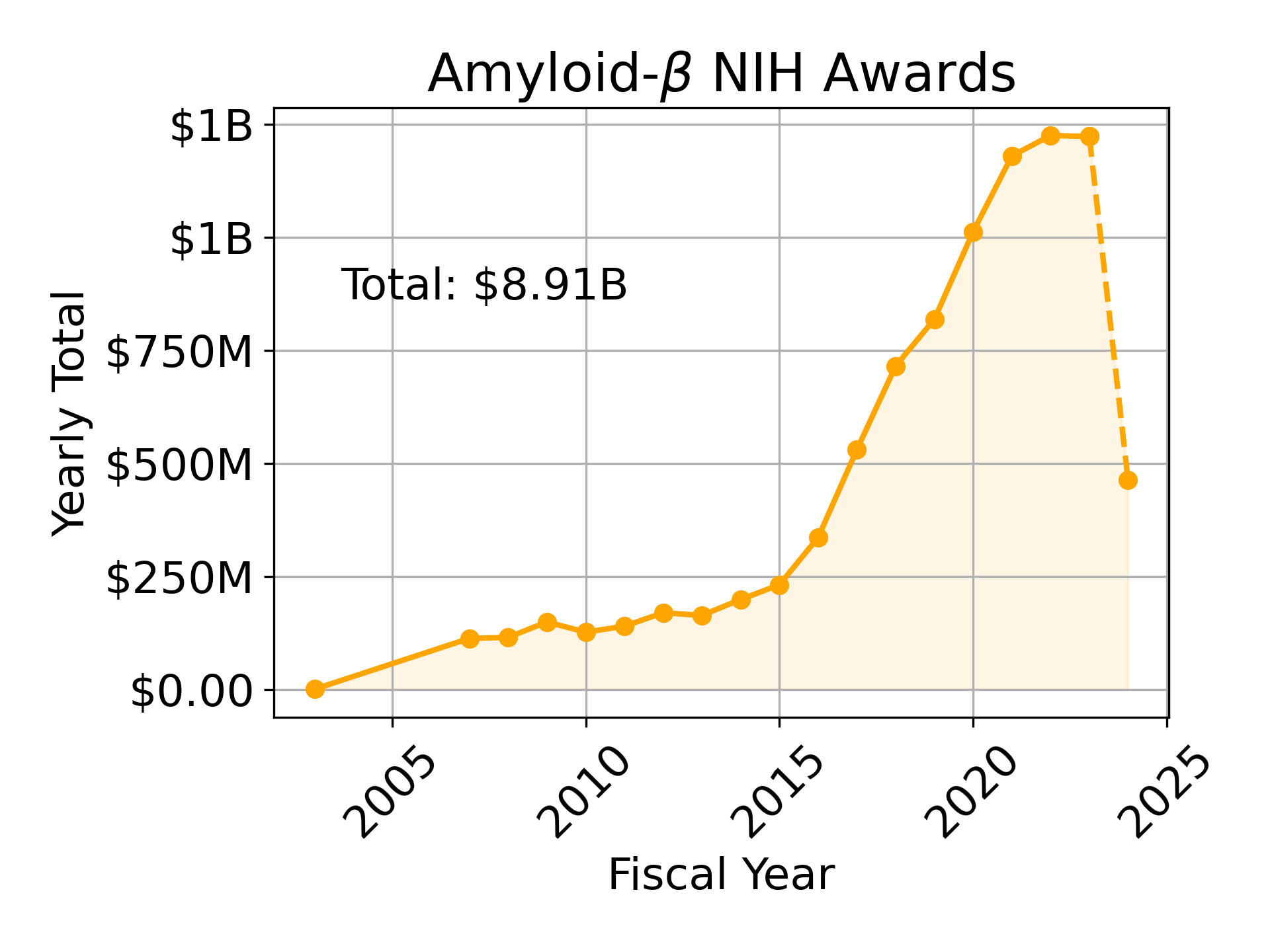}
    \includegraphics[width=0.45\textwidth]{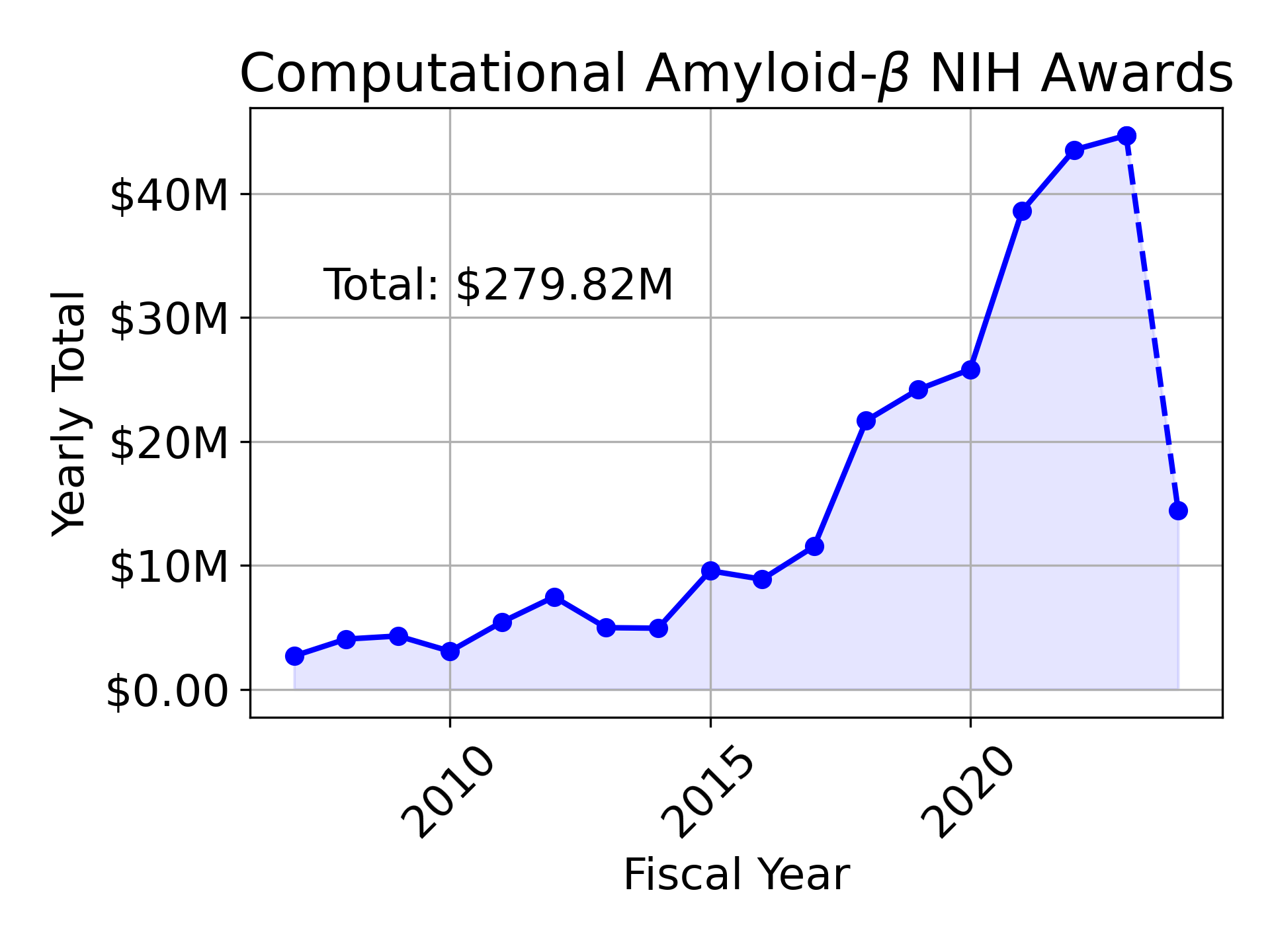}
    \caption{Funded NIH awards per fiscal year. Top: All awards mentioning amyloid-$\beta$ and variants ($\beta$-amyloid, ect). Bottom: Amyloid-$\beta$ awards with ``computational'' in the abstract.}
    \label{fig:enter-label}
\end{figure}
\begin{figure*}[t]
    \centering
    \includegraphics[width=\textwidth]{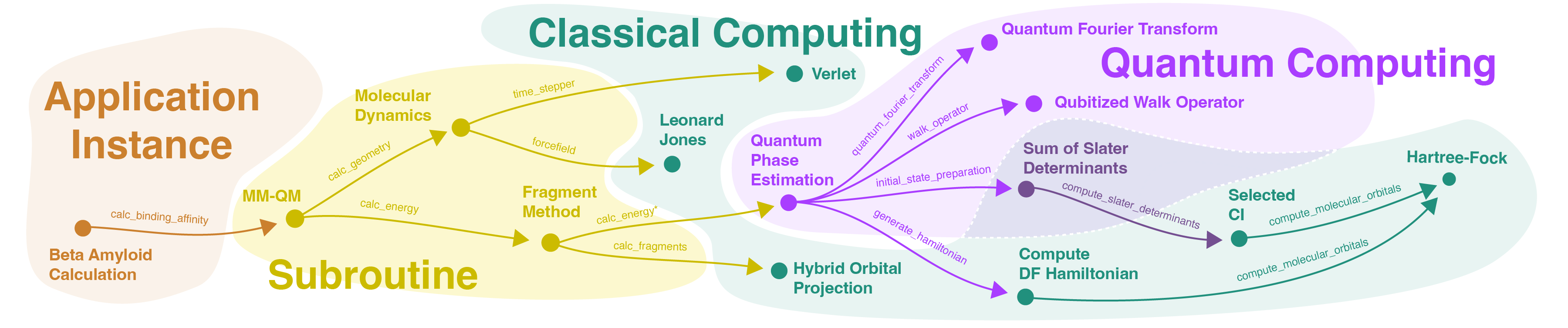}
    \caption{Quantum benchmarking graph (QBG) of the proposed computational workflow.}
    \label{fig:qbg}
\end{figure*}

\section{Workflow}
For this application, the goal is to understand how A$\beta$ interacts with various metal ions. While there
are many ways to attack this problem, one workflow, following the quantum mechanical/molecular mechanics 
(QM/MM) approach is as
follows~\cite{azimi2012binding}. First, an experimental or previously studied geometry is taken from a database,
such as the protein database. Specific choices for this geometry (such as the AB16 protein) are discussed
elsewhere. This structure is then reoptimized at a coarse level (that of, say, 
classical force fields) using geometry optimization techniques. This level of optimization results in several
local minima of similar quality, which are then chosen to be analyzed at more detailed level of theory, here
using a QM/MM approach. The QM/MM performs additional optimization, where a small region is treated fully
quantum mechanically (at, say, a hybrid functional DFT level), and the rest of the protein is treated
via molecular mechanics. The various geometries found from this additional optimization are then analyzed at
an even more accurate level of theory. Here, we use
the fragment molecular orbital (FMO)~\cite{FMO_original} method to divide the AB16 protein into 
multiple smaller fragments. The ground state energy of these fragments is then found directly using the most accurate level of
theory (either a full configuration interaction (FCI)-like algorithm on classical computers or a 
quantum phase estimation (QPE) algorithm on quantum computers). The energies of each fragment are combined
via the FMO algorithm to get the overall energy for the various conformations (geometries).
These energies are compared to determine which conformations are most energetically favorable.
The metal binding affinity can then be calculated by comparing the energy of the protein (which would
be calculated via a similar QM/MM technique) with and without the
additional metal ions. Several corrections to the energy, such as dispersion and solvation, may 
be applied. This workflow is show diagrammatically in Fig.~\ref{fig:qbg}.

The accuracy afforded at the lowest level of the calculation, at the individual fragment level, is necessary to
distinguish between multiple possible structures. QM/MM techniques have been applied to the A$\beta$ system 
before, using much less accurate techniques compared with the proposed FMO + FCI or QPE method. One study, using 
such less-accurate techniques, found 8 structures, all lying within about 40 kJ/mol
of each other, which, due to an expected accuracy of only \~ 20 kJ/mol, had to all be considered equally 
probable~\cite{azimi2012binding}. Overall, the energy of dozens of geometries, both with and without 
an additional metal ion, need to be calculated in an optimization / dynamical loop, resulting in hundreds
of energy evaluations each. 

\subsection{Specific Hamiltonian}
We use the AB16 protein (PDB ID:1ZE9)~\cite{1ZE9} as our protein of interest. The choice for the specific 
protein structure is motivated by the fact that it is the minimal metal coordination domain within the larger AB42 
structure. With a structure identified at physiologically relevant pH by solution NMR, AB16 is realistic and 
biologically relevant starting point. Furthermore, this structure has been previously studied using 
classical techniques, showcasing interest from the community as well as previous results to benchmark against. 
The atomic structure of AB16 is shown in Fig.~\ref{fig:protein}. As part of our workflow, 
we divide the protein into individual fragments which are solved separately and whose solutions are combined
classically. The most interesting, and likely most difficult, fragments to solve are the ones
which involve the metal ion. One such potential binding site, His6, with a copper ion, is shown
in Fig.~\ref{fig:protein}. Our fragmentation scheme results in 15 fragments.

Given the positions of the atoms, we represent our Hamiltonian using Gaussian type orbitals (GTO).
To provide a wide range of resource estimates, we use several basis sets, including, in order
of increasing numbers of basis functions, STO-3G, 6-31g* and cc-pvdz. 
The largest basis sets (Dunning's correlation consistent basis sets~\cite{dunning1989gaussian}) 
studied here are consistent with those that are used for highly-accurate 
quantum chemistry studies. No active space is used; we directly correlate all electrons.

\subsection{Specific Algorithm Descriptions}
\subsubsection{Fragment Molecular Orbital}
The first step in the pipeline is generating fragments from the larger protein sequence. This is done using 
Fragment Molecular Orbital (FMO) method \cite{FMO1}. This method preserves chemical information by breaking bonds
heterolytically and moving a bond with a proton. The method follows an energy decomposition approach, where the 
energy of the entire molecule is constructed by summing over monomer fragment energies, dimer fragment energies,
and so on. Overlapping fragment energies are then subtracted. At the monomer step, each monomer’s electronic 
density relaxes with respect to the electric
field of all other monomers, and are recalculated until a self consistent cycle converges. At the dimer step, 
dimer energies are calculated in the field of the monomer densities (which are not recalculated) \cite{FMO2}. 

\subsubsection{Quantum Phase Estimation with Double-Factorized Qubitization}
Quantum phase estimation (QPE) is one of the core algorithms for solving quantum chemistry problems 
on quantum computers~\cite{kitaev1995quantum}. 
The QPE algorithm to estimate the eigenvalue for a unitary $U$ can be characterized by the following key steps:
\begin{itemize}
\item Initialization.
The algorithm starts with two registers. The ancilla register is initialized to $\bigotimes |+\rangle$ state. The data register is prepared in a state that has suitable overlap with the desired eigenstate of the unitary operator $U$.
\item Controlled Unitary Operations.  A series of controlled unitary operations $U^{2k}$ with integer $k$ are applied to the data register, conditioned on the state of the ancilla qubits.  By varying $k$ for each operation, different powers of the phase are encoded (phase kick-back) into the ancilla qubits.  Implementation of controlled unitaries in this step constitutes the major contribution to resource estimation.
\item Inverse Quantum Fourier Transform (QFT). 
The ancilla register now is in a superposition state encoding the eigenvalue of U.  The inverse Quantum Fourier Transform is then applied to this register and upon measuring the ancilla register, one get a bit string that is an estimate of the phase of the eigenvalue of U.  The desired accuracy of the phase estimate determines the number of ancilla qubits needed and hence is a key factor for resource estimation.
\end{itemize}

For QPE to be applied for ground state energy estimation, an initial state with suitable overlap with the ground state is prepared and evolved in time under the action of the Hamiltonian, $H$. Various methods
of time-evolution can be used, including Trotterization~\cite{PhysRevA.64.022319,PhysRevA.91.022311} and qubitization~\cite{Low2019hamiltonian}.
We use the double-factorized qubitization algorithm of Ref.~\cite{von2021quantum}.
In qubitization, the time evolution is performed not through the direct Hamiltonian, $H$, 
but instead through a walk operator, $W=e^{i \sin^{-1}(H)}$~\cite{Low2019hamiltonian}. Compared with the Trotterization of the time-evolution unitary $\exp{[-itH]}$, qubitization provides
a considerable reduction in gate depth at the cost of additional logical qubits.
The technique of qubitization holds great promises for molecular systems in terms of the T-gate complexity.\cite{Babbush_LinearTComplexity_2018, Berry2019qubitizationof}
For our specific amyloid-$\beta$ application, we use the standard 
quantum chemistry electronic Hamiltonian,
\begin{eqnarray}\label{eq:ham}
    H &=& \sum_{ij,\sigma} h_{ij} a^\dagger_{(i,\sigma)} a _{(i,\sigma)} \nonumber\\
    && + \frac{1}{2}\sum_{ijkl,\sigma\rho} h_{ijkl} a^\dagger_{(i,\sigma)} a^\dagger_{(k,\rho)}a_{(l,\rho)} a_{(j,\sigma)},
\end{eqnarray}
where $h_{ij}$ and $h_{ijkl}$ are the one- and two-electron integrals (computed 
via a standard quantum chemistry package, e.g., pyscf~\cite{sun2018pyscf}); $\sigma$ and $\rho$ index 
spin; and $a_{p}$ are Fermion raising and lowering operators. The Hamiltonian, eq.~\eqref{eq:ham}, is decomposed through the so-called double-factorization procedure,
\begin{eqnarray}
    \label{eq:df-ham}
H_{DF} &=& \sum_{ij,\sigma} \bar{h}_{ij} a^\dagger_{i,\sigma} a_{j,\sigma}+\nonumber\\
&&\frac{1}{2} \sum_{r\in [R]} \Big(\sum_{ij,\sigma} \sum_{m \in [M^{(r)}]}
\lambda_m^{(r)}
\vec{R}_{m,i}^{(r)}
\vec{R}_{m,j}^{(r)}
a^\dagger_{i,\sigma}
a_{j,\sigma}
\Big)^2. \nonumber\\
\end{eqnarray}
This expression is derived from Eq.~\ref{eq:ham} through a two-step factorization of the two-electron tensor terms.  We refer the readers to Ref.~\cite{von2021quantum} for the derivation and more details.

Using Majorana representation of fermion operators, the double-factored Hamiltonian $H_{DF}$ is mapped into a sum of squares of one-body Hamiltonians, and the walk operator can then be synthesized.

\section{Resource Estimates}
\begin{figure}
    \centering
    \includegraphics[width=0.45\textwidth]{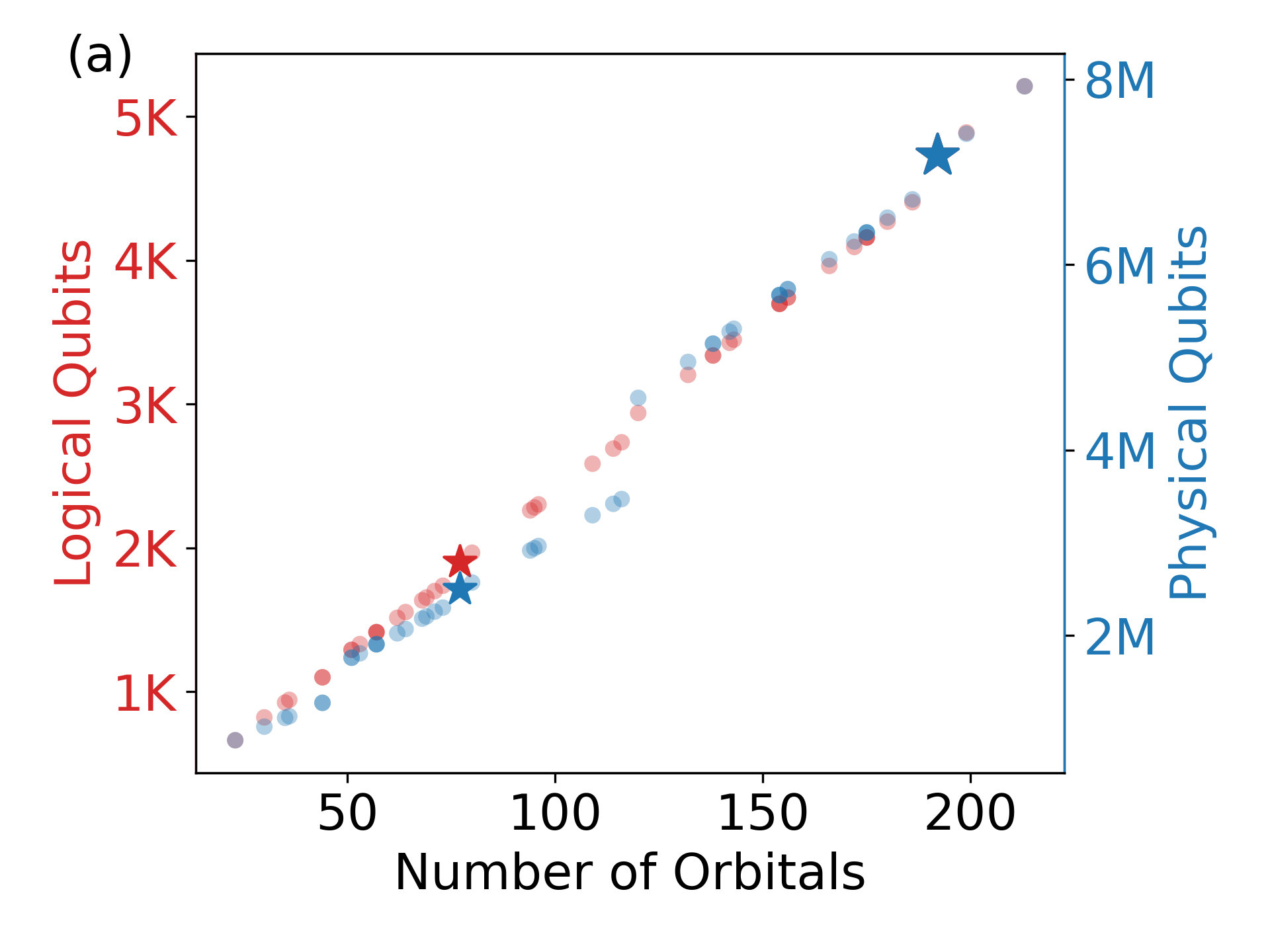}
    \includegraphics[width=0.45\textwidth]{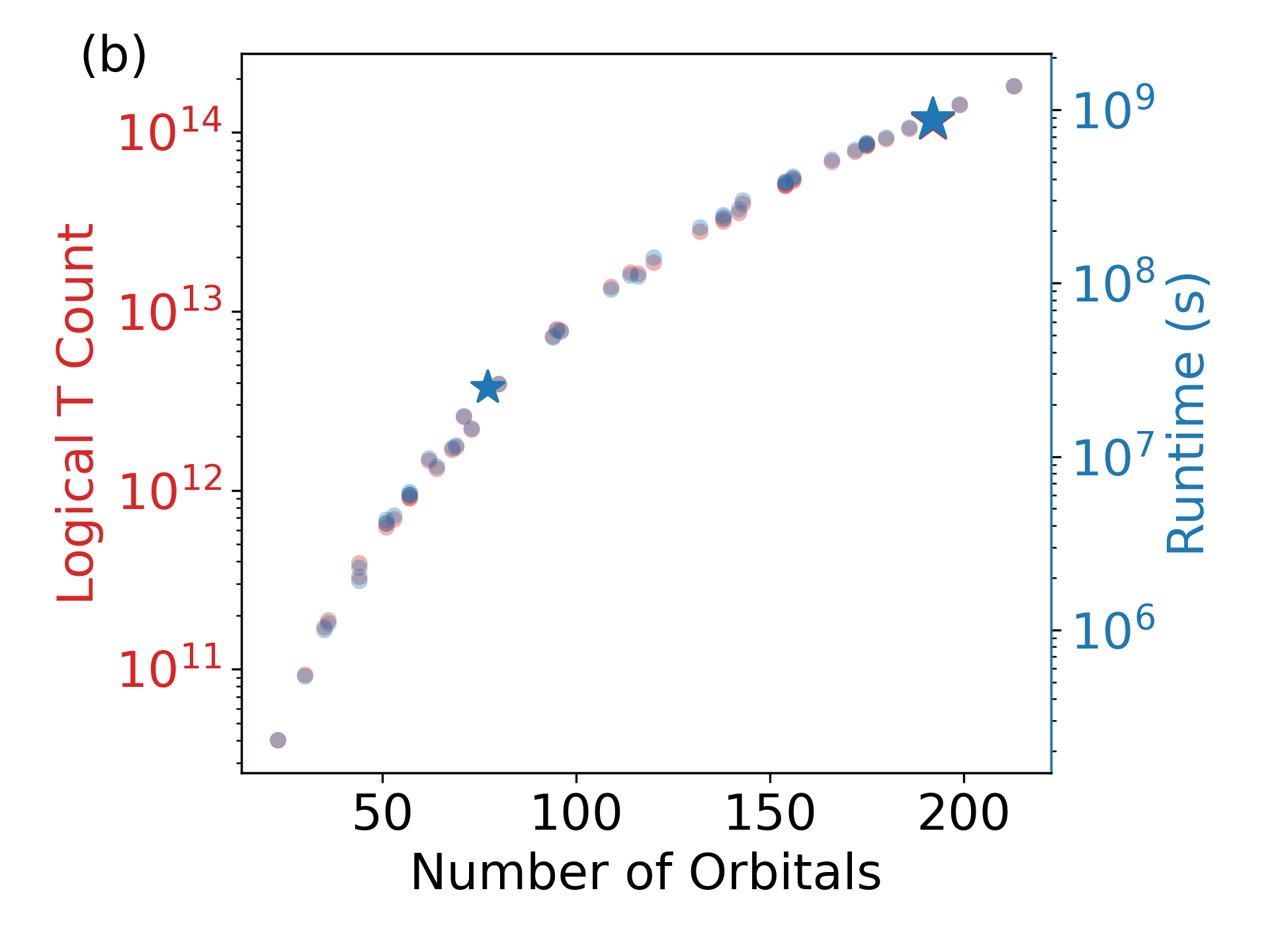}
    \caption{Resource estimates for all fragments of AB16 using various basis sets. Stars represent resource estimates that include a metal center, where we expect we need the accuracy provided by a quantum computer. (a) Number of physical and logical qubits. The step in the number of physical qubits is due to an increase in the surface code distance, $d$. (b) Logical T count and total runtime. A table of all estimates, including other information like surface code distance, number of T factories, etc, can be found in the appendix.}
    \label{fig:all_re}
\end{figure}

We estimate the quantum resources required to solve for the ground state energy of each fragment using the Azure Quantum
Resource Estimator~\cite{beverland2022assessing,Azure_Quantum_Resource_Estimator} implemented in the Azure Quantum
Development Kit~\cite{Microsoft_Azure_Quantum_Development}. For logical resource estimates, we use an accuracy cutoff
of 1mHa, slightly below chemical accuracy for the double-factorized qubitization algorithm. 
For physical resource estimates, we use a model consistent with error rates and gate times for 
optimistic superconducting qubits implementing 
a surface code with a total error budget of 1\%.

\subsubsection{Resource Estimation Details}
We use the Azure Quantum Resource Estimator (AzureQRE)~\cite{Azure_Quantum_Resource_Estimator} to provide
both the logical and physical resource estimates. We briefly describe several
important features of the Azure QRE here; more details can be found in 
Ref.~\cite{beverland2022assessing}. The AzureQRE takes the definition of a logical circuit and compiles it into a
Quantum Intermediate Representation. At the physical level, 
it assumes a 2D nearest-neighbor layout
that has the ability to perform parallel operations. Because the qubits are inherently
noisy, we must estimate the overhead of quantum error correction; specifically,
we estimate the overhead using the surface code~\cite{fowler2012surface}. The distance,
$d$, 
of the surface code parameterizes the level of error suppression and is adjusted
based on the logical depth and physical qubit parameters. Logical qubit movement
and multi-qubit measurements are assumed to be performed via lattice surgery operations. 
The cost to implement T gates is estimated via the use of T state distillation
factories~\cite{bravyi2005universal}. We use physical qubit parameters consistent with an 
optimistic superconducting qubit device, with 50 ns gate times, 100 ns measurement times, and 
$10^{-4}$ Clifford and non-Clifford error rates.
Within the AzureQRE, this has the name \texttt{qubit\_gate\_ns\_e4}.

\subsubsection{Resource Estimation Results}

Logical resource estimates are shown in Fig.~\ref{fig:all_re} for all 15 fragments over the various basis sets. The specific estimates
for the fragments containing metal ions are shown with stars; these fragments are the ones expected to have the strongest correlation, 
necessitating the accuracy provided by quantum computers. The 
number of logical qubits necessary grows linearly with the number of orbitals; the total number of qubits needed is about an
order of magnitude larger than the number of orbitals used to represent the problem due to the additional ancilla qubits needed to perform the 
double-factorized qubitization algorithm. The number of logical T gates grows clearly as $O(n^5)$, with no clear difference between the 
fragments with and without the metal ion. To calculate the His6 binding site with a copper ion in the 6-31g* basis set, we estimate that 4728 
logical qubits implementing 1.17e14 T gates would be required to perform the double-factorized qubitization algorithm. This system is potentially 
the smallest system of practical interest, with
192 orbitals. The 6-31g* basis set 
is the smallest basis set that has the potential to provide somewhat accurate results, though larger basis sets are likely needed
to lower the basis set error. 

To perform such a deep circuit, quantum error correction will be necessary. Physical resource estimates, utilizing quantum error correction, are 
shown in Fig.~\ref{fig:all_re}. The total number of physical qubits is now in the millions. For the His6 binding site with a copper ion in 
the 6-31g* basis set, 7 million physical qubits would be required, with an estimated runtime of 8.8e8s, a little over 28 years.

\section{Conclusion}
In this paper, we provide a specific example of a  biologically-relevant, computational metallorganic problem: calculating the metal-binding 
affinity of the AB16 protein, which is relevant for the study of A$\beta$'s role in neurodegenerative diseases, such as Alzheimer's disease. 
Through a specific computational workflow, involving QM/MM, FMO, and QPE, we provide detailed quantum resource estimates for solving this problem.
The utility of solving this computational problem is at least \$91 million, as evidenced by the NIH funding reports. This protein is only one of the many proteins which contain
a metal ion cofactor. It is expected that a device which could solve the problems discussed
here could likely provide interesting insights into other metalloproteins, such as
FeMoco~\cite{reiher2017elucidating,li2019electronic}. The smallest 
problem of practical interest, that of the His6 binding site interacting with a copper ion, would require 7 million physical qubits
with a run time of a little over 28 years. This points to the need to reduce overheads across the board, from a more compact chemical description,
to better quantum algorithms, better error correction schemes, and faster physical operations. For the AB16 chemical description, active
spaces may be an area to potentially create a more compact chemical description, though active space selection has to be carefully considered
to maintain accuracy. 

\section*{Acknowledgements}
This material is based upon work supported by the Defense Advanced Research Projects Agency under Contract No. HR001122C0074. Any opinions, findings and conclusions or recommendations expressed in this material are those of the authors and do not necessarily reflect the views of the Defense Advanced Research Projects Agency. This work is supported by Wellcome Leap as part of the Quantum for Bio Program.
ZW acknowledges support from DARPA under IAA 8839, Annex 130 through NASA Academic Mission Services (contract NNA16BD14C).

The authors thank John Carpenter for his support in creating high-resolution figures for this paper.
\bibliography{references}
\clearpage
\begin{table}
        \begin{center}
                \begin{tabular}{cccccccccc}
                        \toprule
                        Fragment & Basis & $N_{\mathrm{orb}}$ & $N_{\mathrm{Q,logical}}$ & $T_{\mathrm{count}}$ & Distance & $N_{\mathrm{Q,physical}}$ & $N_{\mathrm{T factories}}$ & $N_{\mathrm{Q,factory}}$ & Runtime (s) \\
                        \midrule
                        8 & sto-3g & 23 & 661 & 4.00e10 & 15 & 8.68e5 & 15 & 2.40e5 & 2.31e5 \\
                        1 & sto-3g & 30 & 820 & 9.29e10 & 15 & 1.01e6 & 15 & 2.40e5 & 5.38e5 \\
                        7 & sto-3g & 35 & 923 & 1.72e11 & 15 & 1.11e6 & 15 & 2.40e5 & 9.97e5 \\
                        0 & sto-3g & 36 & 942 & 1.87e11 & 15 & 1.13e6 & 15 & 2.40e5 & 1.09e6 \\
                        6 & sto-3g & 44 & 1099 & 3.90e11 & 15 & 1.27e6 & 15 & 2.40e5 & 2.28e6 \\
                        11 & sto-3g & 44 & 1099 & 3.28e11 & 15 & 1.27e6 & 15 & 2.40e5 & 1.92e6 \\
                        2 & sto-3g & 51 & 1290 & 6.20e11 & 17 & 1.76e6 & 13 & 2.08e5 & 4.11e6 \\
                        10 & sto-3g & 51 & 1290 & 6.51e11 & 17 & 1.76e6 & 13 & 2.08e5 & 4.31e6 \\
                        14 & sto-3g & 53 & 1330 & 6.87e11 & 17 & 1.81e6 & 13 & 2.08e5 & 4.56e6 \\
                        5 & sto-3g & 57 & 1413 & 9.37e11 & 17 & 1.90e6 & 13 & 2.08e5 & 6.23e6 \\
                        12 & sto-3g & 57 & 1413 & 9.01e11 & 17 & 1.90e6 & 13 & 2.08e5 & 5.99e6 \\
                        13 & sto-3g & 57 & 1413 & 9.06e11 & 17 & 1.90e6 & 13 & 2.08e5 & 6.02e6 \\
                        8 & 6-31g* & 62 & 1513 & 1.46e12 & 17 & 2.02e6 & 13 & 2.08e5 & 9.74e6 \\
                        3 & sto-3g & 64 & 1553 & 1.32e12 & 17 & 2.07e6 & 13 & 2.08e5 & 8.76e6 \\
                        4 & sto-3g & 68 & 1634 & 1.68e12 & 17 & 2.18e6 & 14 & 2.24e5 & 1.12e7 \\
                        9 & sto-3g & 69 & 1654 & 1.74e12 & 17 & 2.20e6 & 14 & 2.24e5 & 1.16e7 \\
                        8 & cc-pvdz & 71 & 1699 & 2.56e12 & 17 & 2.26e6 & 14 & 2.24e5 & 1.71e7 \\
                        15 & sto-3g & 73 & 1736 & 2.18e12 & 17 & 2.30e6 & 14 & 2.24e5 & 1.45e7 \\
                        5+Cu & sto-3g & 77 & 1903 & 3.75e12 & 17 & 2.50e6 & 14 & 2.24e5 & 2.50e7 \\
                        1 & 6-31g* & 80 & 1966 & 3.92e12 & 17 & 2.57e6 & 14 & 2.24e5 & 2.62e7 \\
                        7 & 6-31g* & 94 & 2260 & 7.24e12 & 17 & 2.92e6 & 14 & 2.24e5 & 4.86e7 \\
                        1 & cc-pvdz & 95 & 2281 & 7.95e12 & 17 & 2.94e6 & 14 & 2.24e5 & 5.33e7 \\
                        0 & 6-31g* & 96 & 2302 & 7.82e12 & 17 & 2.96e6 & 14 & 2.24e5 & 5.25e7 \\
                        7 & cc-pvdz & 109 & 2585 & 1.37e13 & 17 & 3.30e6 & 14 & 2.24e5 & 9.19e7 \\
                        0 & cc-pvdz & 114 & 2690 & 1.65e13 & 17 & 3.42e6 & 14 & 2.24e5 & 1.11e8 \\
                        11 & 6-31g* & 116 & 2734 & 1.62e13 & 17 & 3.47e6 & 14 & 2.24e5 & 1.09e8 \\
                        6 & 6-31g* & 120 & 2938 & 1.87e13 & 19 & 4.56e6 & 13 & 2.08e5 & 1.41e8 \\
                        6 & cc-pvdz & 132 & 3203 & 2.78e13 & 19 & 4.95e6 & 13 & 2.08e5 & 2.09e8 \\
                        10 & 6-31g* & 138 & 3338 & 3.28e13 & 19 & 5.15e6 & 13 & 2.08e5 & 2.47e8 \\
                        2 & 6-31g* & 138 & 3338 & 3.18e13 & 19 & 5.15e6 & 13 & 2.08e5 & 2.39e8 \\
                        14 & 6-31g* & 142 & 3426 & 3.55e13 & 19 & 5.28e6 & 13 & 2.08e5 & 2.67e8 \\
                        11 & cc-pvdz & 143 & 3448 & 3.98e13 & 19 & 5.31e6 & 13 & 2.08e5 & 3.00e8 \\
                        13 & 6-31g* & 154 & 3697 & 5.06e13 & 19 & 5.67e6 & 13 & 2.08e5 & 3.81e8 \\
                        5 & 6-31g* & 154 & 3697 & 5.12e13 & 19 & 5.67e6 & 13 & 2.08e5 & 3.86e8 \\
                        12 & 6-31g* & 154 & 3697 & 5.01e13 & 19 & 5.67e6 & 13 & 2.08e5 & 3.77e8 \\
                        10 & cc-pvdz & 156 & 3741 & 5.46e13 & 19 & 5.74e6 & 13 & 2.08e5 & 4.12e8 \\
                        2 & cc-pvdz & 156 & 3741 & 5.34e13 & 19 & 5.74e6 & 13 & 2.08e5 & 4.02e8 \\
                        14 & cc-pvdz & 166 & 3961 & 6.84e13 & 19 & 6.06e6 & 13 & 2.08e5 & 5.16e8 \\
                        3 & 6-31g* & 172 & 4093 & 7.78e13 & 19 & 6.25e6 & 13 & 2.08e5 & 5.87e8 \\
                        13 & cc-pvdz & 175 & 4159 & 8.44e13 & 19 & 6.35e6 & 13 & 2.08e5 & 6.36e8 \\
                        5 & cc-pvdz & 175 & 4159 & 8.56e13 & 19 & 6.35e6 & 13 & 2.08e5 & 6.46e8 \\
                        12 & cc-pvdz & 175 & 4159 & 8.40e13 & 19 & 6.35e6 & 13 & 2.08e5 & 6.34e8 \\
                        4 & 6-31g* & 180 & 4269 & 9.17e13 & 19 & 6.51e6 & 13 & 2.08e5 & 6.91e8 \\
                        9 & 6-31g* & 186 & 4404 & 1.05e14 & 19 & 6.70e6 & 13 & 2.08e5 & 7.90e8 \\
                        5+Cu & 6-31g* & 192 & 4728 & 1.17e14 & 19 & 7.18e6 & 13 & 2.08e5 & 8.83e8 \\
                        3 & cc-pvdz & 199 & 4889 & 1.42e14 & 19 & 7.41e6 & 13 & 2.08e5 & 1.07e9 \\
                        9 & cc-pvdz & 213 & 5211 & 1.81e14 & 19 & 7.92e6 & 14 & 2.51e5 & 1.37e9 \\
                        \bottomrule
                \end{tabular}
        \end{center}
        \caption{Resource estimates for all fragments.}
        \label{tab:resource_estimates}
\end{table}

\begin{table}
        \begin{center}
                \begin{tabular}{cccccccccc}
                        \toprule
                        Fragment & Basis & $N_{\mathrm{orb}}$ & $N_{\mathrm{Q,logical}}$ & $T_{\mathrm{count}}$ & Distance & $N_{\mathrm{Q,physical}}$ & $N_{\mathrm{T factories}}$ & $N_{\mathrm{Q,factory}}$ & Runtime (s) \\
                        \midrule
                        8 & sto-3g & 23 & 661 & 4.00e10 & 15 & 8.68e5 & 15 & 2.40e5 & 2.31e5 \\
                        5 & sto-3g & 57 & 1413 & 9.37e11 & 17 & 1.90e6 & 13 & 2.08e5 & 6.23e6 \\
                        5+Cu & sto-3g & 77 & 1903 & 3.75e12 & 17 & 2.50e6 & 14 & 2.24e5 & 2.50e7 \\
                        1 & 6-31g* & 80 & 1966 & 3.92e12 & 17 & 2.57e6 & 14 & 2.24e5 & 2.62e7 \\
                        7 & 6-31g* & 94 & 2260 & 7.24e12 & 17 & 2.92e6 & 14 & 2.24e5 & 4.86e7 \\
                        1 & cc-pvdz & 95 & 2281 & 7.95e12 & 17 & 2.94e6 & 14 & 2.24e5 & 5.33e7 \\
                        14 & 6-31g* & 142 & 3426 & 3.55e13 & 19 & 5.28e6 & 13 & 2.08e5 & 2.67e8 \\
                        11 & cc-pvdz & 143 & 3448 & 3.98e13 & 19 & 5.31e6 & 13 & 2.08e5 & 3.00e8 \\
                        5 & 6-31g* & 154 & 3697 & 5.12e13 & 19 & 5.67e6 & 13 & 2.08e5 & 3.86e8 \\
                        5 & cc-pvdz & 175 & 4159 & 8.56e13 & 19 & 6.35e6 & 13 & 2.08e5 & 6.46e8 \\
                        12 & cc-pvdz & 175 & 4159 & 8.40e13 & 19 & 6.35e6 & 13 & 2.08e5 & 6.34e8 \\
                        9 & 6-31g* & 186 & 4404 & 1.05e14 & 19 & 6.70e6 & 13 & 2.08e5 & 7.90e8 \\
                        5+Cu & 6-31g* & 192 & 4728 & 1.17e14 & 19 & 7.18e6 & 13 & 2.08e5 & 8.83e8 \\
                        3 & cc-pvdz & 199 & 4889 & 1.42e14 & 19 & 7.41e6 & 13 & 2.08e5 & 1.07e9 \\
                        9 & cc-pvdz & 213 & 5211 & 1.81e14 & 19 & 7.92e6 & 14 & 2.51e5 & 1.37e9 \\
                        \bottomrule
                \end{tabular}
        \end{center}
        \caption{Resource estimates for selected fragments.}
        \label{tab:resource_estimates}
\end{table}

\begin{table}
        \begin{center}
                \begin{tabular}{llll}
                        \toprule
                        \multicolumn{4}{c}{Fragment 13} \\
                        \toprule
                        C & -3.462 & -4.282 & 1.523 \\
                        O & -2.969 & -4.629 & 0.451 \\
                        N & -4.759 & -4.103 & 1.729 \\
                        H & -5.144 & -3.768 & 2.588 \\
                        C & -5.729 & -4.392 & 0.687 \\
                        H & -5.528 & -5.408 & 0.346 \\
                        C & -7.153 & -4.362 & 1.247 \\
                        H & -7.841 & -4.835 & 0.513 \\
                        H & -7.824 & -4.836 & 0.531 \\
                        N & -8.419 & -2.681 & 2.673 \\
                        H & -8.697 & -3.335 & 3.378 \\
                        C & -7.664 & -2.973 & 1.551 \\
                        C & -8.711 & -1.389 & 2.656 \\
                        H & -9.294 & -0.854 & 3.406 \\
                        N & -8.153 & -0.846 & 1.536 \\
                        C & -7.519 & -1.801 & 0.868 \\
                        H & -6.976 & -1.672 & -0.068 \\
                        \bottomrule
                \end{tabular}
        \end{center}
        \caption{Geometry for fragment 13.}
        \label{tab:fragment_13}
\end{table}
\begin{table}
        \begin{center}
                \begin{tabular}{llll}
                        \toprule
                        \multicolumn{4}{c}{Fragment 4} \\
                        \toprule
                        C & -8.802 & 4.408 & -4.484 \\
                        O & -8.920 & 3.816 & -3.413 \\
                        N & -7.956 & 5.403 & -4.702 \\
                        H & -7.906 & 5.928 & -5.550 \\
                        C & -7.006 & 5.815 & -3.683 \\
                        H & -6.821 & 4.951 & -3.045 \\
                        C & -5.752 & 6.305 & -4.411 \\
                        H & -6.053 & 6.812 & -5.354 \\
                        H & -6.054 & 6.856 & -5.302 \\
                        C & -4.793 & 5.186 & -4.822 \\
                        C & -4.072 & 4.527 & -3.875 \\
                        H & -4.177 & 4.797 & -2.824 \\
                        C & -3.182 & 3.488 & -4.256 \\
                        H & -2.604 & 2.960 & -3.498 \\
                        C & -3.051 & 3.152 & -5.568 \\
                        H & -2.368 & 2.354 & -5.860 \\
                        C & -4.663 & 4.849 & -6.133 \\
                        H & -5.241 & 5.377 & -6.891 \\
                        C & -3.773 & 3.811 & -6.514 \\
                        H & -3.668 & 3.541 & -7.565 \\
                        \bottomrule
                \end{tabular}
        \end{center}
        \caption{Geometry for fragment 4.}
        \label{tab:fragment_4}
\end{table}
\begin{table}
        \begin{center}
                \begin{tabular}{llll}
                        \toprule
                        \multicolumn{4}{c}{Fragment 7} \\
                        \toprule
                        C & -5.445 & 6.487 & 1.509 \\
                        O & -4.556 & 5.789 & 1.995 \\
                        N & -5.237 & 7.632 & 0.875 \\
                        H & -5.940 & 8.227 & 0.494 \\
                        C & -3.892 & 8.142 & 0.672 \\
                        H & -3.339 & 7.319 & 0.220 \\
                        C & -3.898 & 9.362 & -0.251 \\
                        H & -2.922 & 9.882 & -0.175 \\
                        H & -2.919 & 9.838 & -0.206 \\
                        C & -4.966 & 10.409 & 0.070 \\
                        O & -5.599 & 10.266 & 1.139 \\
                        O & -5.126 & 11.329 & -0.761 \\
                        \bottomrule
                \end{tabular}
        \end{center}
        \caption{Geometry for fragment 7.}
        \label{tab:fragment_7}
\end{table}
\begin{table}
        \begin{center}
                \begin{tabular}{llll}
                        \toprule
                        \multicolumn{4}{c}{Fragment 9} \\
                        \toprule
                        C & -3.74 & 8.105 & 5.22 \\
                        O & -3.733 & 8.271 & 6.439 \\
                        N & -3.709 & 6.922 & 4.624 \\
                        H & -3.712 & 6.807 & 3.631 \\
                        C & -3.668 & 5.694 & 5.399 \\
                        H & -4.566 & 5.655 & 6 \\
                        H & -2.780 & 5.688 & 6.031 \\
                        \bottomrule
                \end{tabular}
        \end{center}
        \caption{Geometry for fragment 9.}
        \label{tab:fragment_9}
\end{table}
\begin{table}
        \begin{center}
                \begin{tabular}{llll}
                        \toprule
                        \multicolumn{4}{c}{Fragment 8} \\
                        \toprule
                        C & -3.305 & 8.571 & 2.019 \\
                        O & -2.087 & 8.614 & 2.184 \\
                        N & -4.200 & 8.877 & 2.947 \\
                        H & -5.191 & 8.841 & 2.819 \\
                        C & -3.786 & 9.305 & 4.273 \\
                        H & -2.790 & 9.728 & 4.141 \\
                        C & -4.727 & 10.380 & 4.823 \\
                        H & -5.710 & 9.902 & 5.037 \\
                        H & -5.685 & 9.927 & 5.077 \\
                        O & -4.934 & 11.436 & 3.889 \\
                        H & -5.692 & 11.946 & 4.201 \\
                        \bottomrule
                \end{tabular}
        \end{center}
        \caption{Geometry for fragment 8.}
        \label{tab:fragment_8}
\end{table}
\begin{table}
        \begin{center}
                \begin{tabular}{llll}
                        \toprule
                        \multicolumn{4}{c}{Fragment 1} \\
                        \toprule
                        N & -8.329 & 9.458 & -6.956 \\
                        H & -7.576 & 10.173 & -6.893 \\
                        H & -9.011 & 9.737 & -7.690 \\
                        H & -7.912 & 8.537 & -7.202 \\
                        C & -9.009 & 9.356 & -5.676 \\
                        H & -8.275 & 9.055 & -4.929 \\
                        C & -9.626 & 10.731 & -5.416 \\
                        H & -10.268 & 11.011 & -6.275 \\
                        H & -10.293 & 10.976 & -6.243 \\
                        C & -8.619 & 11.870 & -5.247 \\
                        O & -7.925 & 12.167 & -6.243 \\
                        O & -8.566 & 12.419 & -4.125 \\
                        \bottomrule
                \end{tabular}
        \end{center}
        \caption{Geometry for fragment 1.}
        \label{tab:fragment_1}
\end{table}
\begin{table}
        \begin{center}
                \begin{tabular}{llll}
                        \toprule
                        \multicolumn{4}{c}{Fragment 11} \\
                        \toprule
                        C & -2.331 & 1.78 & 3.794 \\
                        O & -1.704 & 1.695 & 4.849 \\
                        N & -3.036 & 0.794 & 3.259 \\
                        H & -3.519 & 0.843 & 2.389 \\
                        C & -3.161 & -0.488 & 3.933 \\
                        H & -2.434 & -0.461 & 4.744 \\
                        C & -4.563 & -0.662 & 4.522 \\
                        H & -5.130 & -1.364 & 3.868 \\
                        H & -5.139 & -1.348 & 3.901 \\
                        C & -5.288 & 0.681 & 4.617 \\
                        H & -5.969 & 0.672 & 5.487 \\
                        H & -5.908 & 0.701 & 5.513 \\
                        C & -6.160 & 0.922 & 3.383 \\
                        O & -5.568 & 1.133 & 2.302 \\
                        O & -7.399 & 0.891 & 3.548 \\
                        \bottomrule
                \end{tabular}
        \end{center}
        \caption{Geometry for fragment 11.}
        \label{tab:fragment_11}
\end{table}
\begin{table}
        \begin{center}
                \begin{tabular}{llll}
                        \toprule
                        \multicolumn{4}{c}{Fragment 5} \\
                        \toprule
                        C & -7.57 & 6.959 & -2.838 \\
                        O & -7.103 & 8.093 & -2.930 \\
                        N & -8.566 & 6.621 & -2.032 \\
                        H & -8.980 & 5.714 & -2.048 \\
                        C & -9.109 & 7.569 & -1.075 \\
                        H & -8.814 & 8.548 & -1.454 \\
                        C & -10.634 & 7.466 & -1.001 \\
                        H & -10.924 & 6.606 & -1.650 \\
                        H & -10.974 & 6.623 & -1.602 \\
                        C & -11.295 & 8.753 & -1.498 \\
                        H & -11.073 & 8.869 & -2.581 \\
                        H & -11.033 & 8.921 & -2.542 \\
                        C & -10.858 & 9.954 & -0.657 \\
                        H & -11.741 & 10.601 & -0.449 \\
                        H & -11.717 & 10.589 & -0.440 \\
                        N & -9.827 & 10.731 & -1.381 \\
                        H & -10.066 & 11.107 & -2.277 \\
                        C & -8.593 & 10.961 & -0.913 \\
                        N & -8.250 & 10.537 & 0.311 \\
                        H & -8.914 & 10.041 & 0.872 \\
                        H & -7.332 & 10.715 & 0.664 \\
                        N & -7.701 & 11.616 & -1.669 \\
                        H & -7.975 & 11.991 & -2.554 \\
                        H & -6.761 & 11.731 & -1.347 \\
                        \bottomrule
                \end{tabular}
        \end{center}
        \caption{Geometry for fragment 5.}
        \label{tab:fragment_5}
\end{table}
\begin{table}
        \begin{center}
                \begin{tabular}{llll}
                        \toprule
                        \multicolumn{4}{c}{Fragment 16} \\
                        \toprule
                        C & -1.871 & -2.898 & -4.128 \\
                        O & -2.849 & -3.469 & -4.607 \\
                        N & -1.005 & -2.177 & -4.826 \\
                        H & -0.209 & -1.723 & -4.429 \\
                        C & -1.166 & -1.997 & -6.259 \\
                        H & -1.546 & -2.934 & -6.666 \\
                        C & -2.215 & -0.920 & -6.548 \\
                        H & -3.079 & -1.113 & -5.870 \\
                        H & -3.078 & -1.064 & -5.899 \\
                        C & -2.657 & -0.965 & -8.012 \\
                        H & -1.793 & -0.696 & -8.662 \\
                        H & -1.816 & -0.714 & -8.658 \\
                        C & -3.810 & 0.007 & -8.267 \\
                        H & -4.631 & -0.225 & -7.550 \\
                        H & -4.632 & -0.212 & -7.585 \\
                        C & -4.300 & -0.089 & -9.713 \\
                        H & -3.499 & 0.228 & -10.418 \\
                        H & -3.499 & 0.199 & -10.394 \\
                        N & -5.476 & 0.785 & -9.922 \\
                        H & -5.210 & 1.739 & -9.784 \\
                        H & -5.819 & 0.666 & -10.853 \\
                        H & -6.193 & 0.541 & -9.269 \\
                        C & 0.198 & -1.711 & -6.889 \\
                        O & 1.204 & -1.674 & -6.131 \\
                        O & 0.247 & -1.527 & -8.135 \\
                        \bottomrule
                \end{tabular}
        \end{center}
        \caption{Geometry for fragment 16.}
        \label{tab:fragment_16}
\end{table}
\begin{table}
        \begin{center}
                \begin{tabular}{llll}
                        \toprule
                        \multicolumn{4}{c}{Fragment 12} \\
                        \toprule
                        C & -2.825 & -1.629 & 2.97 \\
                        O & -2.380 & -1.389 & 1.849 \\
                        N & -3.051 & -2.845 & 3.444 \\
                        H & -3.547 & -3.042 & 4.286 \\
                        C & -2.572 & -4.023 & 2.740 \\
                        H & -1.560 & -3.811 & 2.396 \\
                        C & -2.507 & -5.214 & 3.698 \\
                        H & -1.845 & -4.947 & 4.521 \\
                        C & -3.886 & -5.522 & 4.284 \\
                        H & -4.559 & -5.837 & 3.486 \\
                        H & -3.798 & -6.319 & 5.022 \\
                        H & -4.285 & -4.627 & 4.763 \\
                        C & -1.920 & -6.445 & 3.003 \\
                        H & -2.562 & -6.732 & 2.170 \\
                        H & -0.923 & -6.210 & 2.630 \\
                        H & -1.858 & -7.268 & 3.714 \\
                        \bottomrule
                \end{tabular}
        \end{center}
        \caption{Geometry for fragment 12.}
        \label{tab:fragment_12}
\end{table}
\begin{table}
        \begin{center}
                \begin{tabular}{llll}
                        \toprule
                        \multicolumn{4}{c}{Fragment 15} \\
                        \toprule
                        C & -3.372 & -1.484 & -2.018 \\
                        O & -2.989 & -0.596 & -2.779 \\
                        N & -2.765 & -2.651 & -1.862 \\
                        H & -3.064 & -3.364 & -1.230 \\
                        C & -1.577 & -2.986 & -2.629 \\
                        H & -0.834 & -2.238 & -2.352 \\
                        C & -1.055 & -4.375 & -2.253 \\
                        H & -0.111 & -4.530 & -2.825 \\
                        H & -0.124 & -4.573 & -2.783 \\
                        C & -2.082 & -5.457 & -2.593 \\
                        H & -3.066 & -5.215 & -2.137 \\
                        H & -3.053 & -5.184 & -2.179 \\
                        C & -1.650 & -6.816 & -2.039 \\
                        O & -1.139 & -7.669 & -2.746 \\
                        N & -1.883 & -6.970 & -0.739 \\
                        H & -2.298 & -6.226 & -0.214 \\
                        H & -1.642 & -7.828 & -0.286 \\
                        \bottomrule
                \end{tabular}
        \end{center}
        \caption{Geometry for fragment 15.}
        \label{tab:fragment_15}
\end{table}
\begin{table}
        \begin{center}
                \begin{tabular}{llll}
                        \toprule
                        \multicolumn{4}{c}{Fragment 10} \\
                        \toprule
                        C & -3.657 & 4.468 & 4.484 \\
                        O & -4.668 & 3.780 & 4.350 \\
                        N & -2.502 & 4.231 & 3.878 \\
                        H & -1.686 & 4.799 & 3.976 \\
                        C & -2.339 & 3.087 & 2.998 \\
                        H & -3.179 & 3.075 & 2.303 \\
                        C & -0.979 & 3.266 & 2.321 \\
                        H & -0.863 & 4.328 & 2.011 \\
                        H & -0.848 & 4.317 & 2.065 \\
                        C & -0.798 & 2.422 & 1.057 \\
                        C & -0.690 & 1.049 & 1.154 \\
                        H & -0.734 & 0.566 & 2.130 \\
                        C & -0.519 & 0.255 & -0.035 \\
                        H & -0.432 & -0.829 & 0.027 \\
                        C & -0.468 & 0.889 & -1.237 \\
                        O & -0.306 & 0.140 & -2.360 \\
                        H & -0.387 & 0.718 & -3.172 \\
                        C & -0.742 & 3.033 & -0.179 \\
                        H & -0.827 & 4.117 & -0.255 \\
                        C & -0.571 & 2.239 & -1.368 \\
                        H & -0.525 & 2.709 & -2.350 \\
                        \bottomrule
                \end{tabular}
        \end{center}
        \caption{Geometry for fragment 10.}
        \label{tab:fragment_10}
\end{table}
\begin{table}
        \begin{center}
                \begin{tabular}{llll}
                        \toprule
                        \multicolumn{4}{c}{Fragment 2} \\
                        \toprule
                        C & -10.126 & 8.311 & -5.73 \\
                        O & -10.623 & 7.875 & -4.693 \\
                        N & -10.487 & 7.941 & -6.950 \\
                        H & -10.072 & 8.303 & -7.784 \\
                        C & -11.539 & 6.958 & -7.152 \\
                        H & -12.480 & 7.410 & -6.839 \\
                        C & -11.631 & 6.608 & -8.639 \\
                        H & -11.935 & 7.491 & -9.202 \\
                        H & -10.657 & 6.269 & -8.993 \\
                        H & -12.366 & 5.816 & -8.781 \\
                        \bottomrule
                \end{tabular}
        \end{center}
        \caption{Geometry for fragment 2.}
        \label{tab:fragment_2}
\end{table}
\begin{table}
        \begin{center}
                \begin{tabular}{llll}
                        \toprule
                        \multicolumn{4}{c}{Fragment 14} \\
                        \toprule
                        C & -5.531 & -3.425 & -0.481 \\
                        O & -6.074 & -3.636 & -1.565 \\
                        N & -4.752 & -2.386 & -0.222 \\
                        H & -4.214 & -2.290 & 0.614 \\
                        C & -4.628 & -1.289 & -1.167 \\
                        H & -5.505 & -1.332 & -1.813 \\
                        C & -4.649 & 0.058 & -0.442 \\
                        H & -5.062 & -0.087 & 0.580 \\
                        H & -5.045 & -0.088 & 0.563 \\
                        N & -6.539 & 1.755 & -0.538 \\
                        H & -6.834 & 1.647 & 0.412 \\
                        C & -5.468 & 1.119 & -1.139 \\
                        C & -7.046 & 2.625 & -1.399 \\
                        H & -7.899 & 3.281 & -1.221 \\
                        N & -6.314 & 2.556 & -2.548 \\
                        C & -5.360 & 1.647 & -2.392 \\
                        H & -4.615 & 1.370 & -3.139 \\
                        \bottomrule
                \end{tabular}
        \end{center}
        \caption{Geometry for fragment 14.}
        \label{tab:fragment_14}
\end{table}
\begin{table}
        \begin{center}
                \begin{tabular}{llll}
                        \toprule
                        \multicolumn{4}{c}{Fragment 6} \\
                        \toprule
                        C & -8.52 & 7.317 & 0.314 \\
                        O & -8.905 & 7.969 & 1.284 \\
                        N & -7.595 & 6.370 & 0.367 \\
                        H & -7.316 & 5.823 & -0.421 \\
                        C & -6.910 & 6.060 & 1.610 \\
                        H & -7.398 & 6.646 & 2.390 \\
                        C & -7.074 & 4.581 & 1.967 \\
                        H & -6.506 & 3.967 & 1.235 \\
                        H & -6.507 & 3.981 & 1.255 \\
                        N & -8.857 & 2.775 & 1.828 \\
                        H & -8.227 & 1.999 & 1.781 \\
                        C & -8.507 & 4.106 & 1.975 \\
                        C & -10.178 & 2.682 & 1.876 \\
                        H & -10.758 & 1.763 & 1.791 \\
                        N & -10.683 & 3.936 & 2.053 \\
                        C & -9.675 & 4.798 & 2.113 \\
                        H & -9.765 & 5.876 & 2.250 \\
                        \bottomrule
                \end{tabular}
        \end{center}
        \caption{Geometry for fragment 6.}
        \label{tab:fragment_6}
\end{table}
\begin{table}
        \begin{center}
                \begin{tabular}{llll}
                        \toprule
                        \multicolumn{4}{c}{Fragment 3} \\
                        \toprule
                        C & -11.26 & 5.732 & -6.281 \\
                        O & -12.119 & 5.306 & -5.509 \\
                        N & -10.057 & 5.199 & -6.432 \\
                        H & -9.351 & 5.548 & -7.042 \\
                        C & -9.663 & 4.016 & -5.687 \\
                        H & -10.595 & 3.566 & -5.344 \\
                        C & -8.928 & 3.020 & -6.586 \\
                        H & -8.894 & 2.038 & -6.061 \\
                        H & -8.883 & 2.047 & -6.098 \\
                        C & -7.511 & 3.507 & -6.899 \\
                        H & -7.557 & 4.521 & -7.335 \\
                        H & -7.548 & 4.534 & -7.263 \\
                        C & -6.843 & 2.612 & -7.945 \\
                        O & -7.085 & 1.388 & -7.883 \\
                        O & -6.105 & 3.174 & -8.783 \\
                        \bottomrule
                \end{tabular}
        \end{center}
        \caption{Geometry for fragment 3.}
        \label{tab:fragment_3}
\end{table}
\end{document}